\documentclass[superscriptaddress,onecolumn,showpacs,prb,floatfix]{revtex4}
\usepackage{graphicx}
\usepackage{tabularx}
\pdfoutput=1

\begin{document}
\title{Evidence for Two Current Conduction in Iron}
\author{I.~A.~Campbell}
\affiliation{Physique des Solides, Facult\'e des Sciences, 91 Orsay, France}
\author{A.~Fert}
\affiliation{Physique des Solides, Facult\'e des Sciences, 91 Orsay, France}
\author{A.~R.~Pomeroy}
\affiliation{Department of Physics, University of Essex}

\begin{abstract}

Measurements of resistivities of dilute iron based alloys show strong deviations from Matthiessen's rule. These deviations can be explained by a model in which spin $\uparrow$ and spin $\downarrow$ electrons conduct in parallel. The results are consistent with the theory of impurity shielding in these alloys.\\

[This paper provided the first experimental demonstration of two current conduction at low temperatures in a ferromagnetic metal. One direct consequence of this property is the Giant Magnetoresistance discovered in 1988 by the groups of Albert Fert and of Peter Gr\"unberg].\\

Published in the PHILOSOPHICAL MAGAZINE, Vol. 15, p. 977, May 1967

\end{abstract}
\maketitle

A number of authors (Mott \cite{Mott1936,Mott1964}, see also Weiss and Marotta \cite{Weiss1959}) have suggested that some anomalies in the resistivities of ferromagnetic metals could be explained by a model in which the spin-up and spin-down electrons carry current in parallel. Measurements which we have made on the resistivities of iron based alloys containing dilute impurities provide evidence that this is indeed the case. There are some interesting consequences if this interpretation is correct.

In order to have a standard of comparison, we compare the resistivities throughout with those which would be expected if Matthiessen's rule was obeyed. This rule (Matthiessen \cite{Matthiessen1864}) follows from the assumption that there is one type of current carrier and that the scattering processes which cause resistance are independent; thus for an alloy at temperature $T$ the resistivity predicted by the rule is:
\begin{equation}	
\rho_{m}(T)= \rho(0)+ \rho_{h}(T)
\end{equation}	
where $\rho(0)$ is the residual resistivity of the alloy, and $\rho_{h}(T)$ is the resistivity of the pure host-metal at temperature $T$. (This temperature dependent term includes all the different scattering terms for the host-metal.) For a metal containing two impurities, $A$ and $B$, it would follow that
\begin{equation}
\rho_{A+B}(0) = \rho_{A}(0) +\rho_{B}(0)
\end{equation}
the total residual resistivity being the sum of the residual resistivities of two samples with the impurities present separately at the same con­centrations as before.

Neither of these rules is obeyed even approximately in iron alloys. Table 1 gives the resistivities at $4 K$ and at room temperature of a number
	of dilute alloy samples. It can be seen that for all the alloys where the impurity is a transition metal there is a deviation from Matthiessen's rule at room temperature; for some the deviation
\begin{equation}
\Delta\rho(T)= \rho(T) - [\rho(0) + \rho_{Fe}(T)]
\end{equation}
at $T = 293 K$ is as large as the resistivity of pure iron at this temperature, and larger than the residual resistivity. Samples where the impurities were $Al$ or $Si$, however, show little or no deviation. Measurements on some of the samples with transition metal impurities over the range $4K$ to $70K$ showed that the deviation $\Delta\rho(T)$ is approximately proportional
to $T^2$ at low temperatures (fig. 1). At higher temperatures the rate of increase slowed down, but $\Delta\rho$ was still increasing at room temperature (fig. 3).

Secondly, when two impurities are present simultaneously the total residual resistivity is found to be greater than the sum of the separate residual resistivities, for certain combinations of impurities. Some examples are given in table 2. ln general we find the following rule to hold: when an impurity giving no temperature deviation as defined above (i.e. $Al$ or $Si$) is present simultaneously with an impurity giving a strong deviation (e.g. $Mn$, $V$ or $Cr$) the residual resistivity of the sample is much higher than would be expected from the sum rule eqn (2), and the sample shows a small temperature deviation. When two impurities having strong temperature deviations are present together the sum rule for the residual resistances holds (at least approximately) and the sample shows strong temperature deviation.

\begin{table}
\caption{The room temperature deviations of single impurity alloys. $\Delta\rho$ is the deviation at room temperature from the resisitivity predicted by Matthiessen's rule.}
\begin{tabular*}{\columnwidth}{c r c r c r c r c}
\hline
\hline
Impurity & & conc. wt $\%$ & & $\rho_{4.2 K}\mu\Omega cm$ & & $\rho_{293 K} \mu\Omega cm$ & & $\Delta\rho \mu\Omega cm$\\
Al & & 3.15& & 33.7 & & 43.9 & & 0.3\\
Si & & 2.91 & & 33.8 & & 43.9 & & 0.1\\
V & &  1.88  & & 2.8  & & 17.0 & & 4.1\\
Mn & & 1.88  & & 3.2 & & 19.6 & & 6.4\\
Mn & & 1.45 & &  2.6 & & 17.4 & & 4.9\\
Ni & & 1.97 & & 3.7 & & 14.8 & & 1.3\\
Ti & &  1.95  & & 5.8 & & 19.0 & & 3.3\\
Os & & 6.5 & & 7.6 & & 21.3 & & 4.6\\
Os & & 3.33 & &  3.9 & & 17.4 & & 3.3\\
Re & & 3.26 & & 3.2 & & 15.9 & & 2.9\\

\hline
\hline
\end{tabular*}
\end{table}

\begin{figure}
\includegraphics[width=4in]{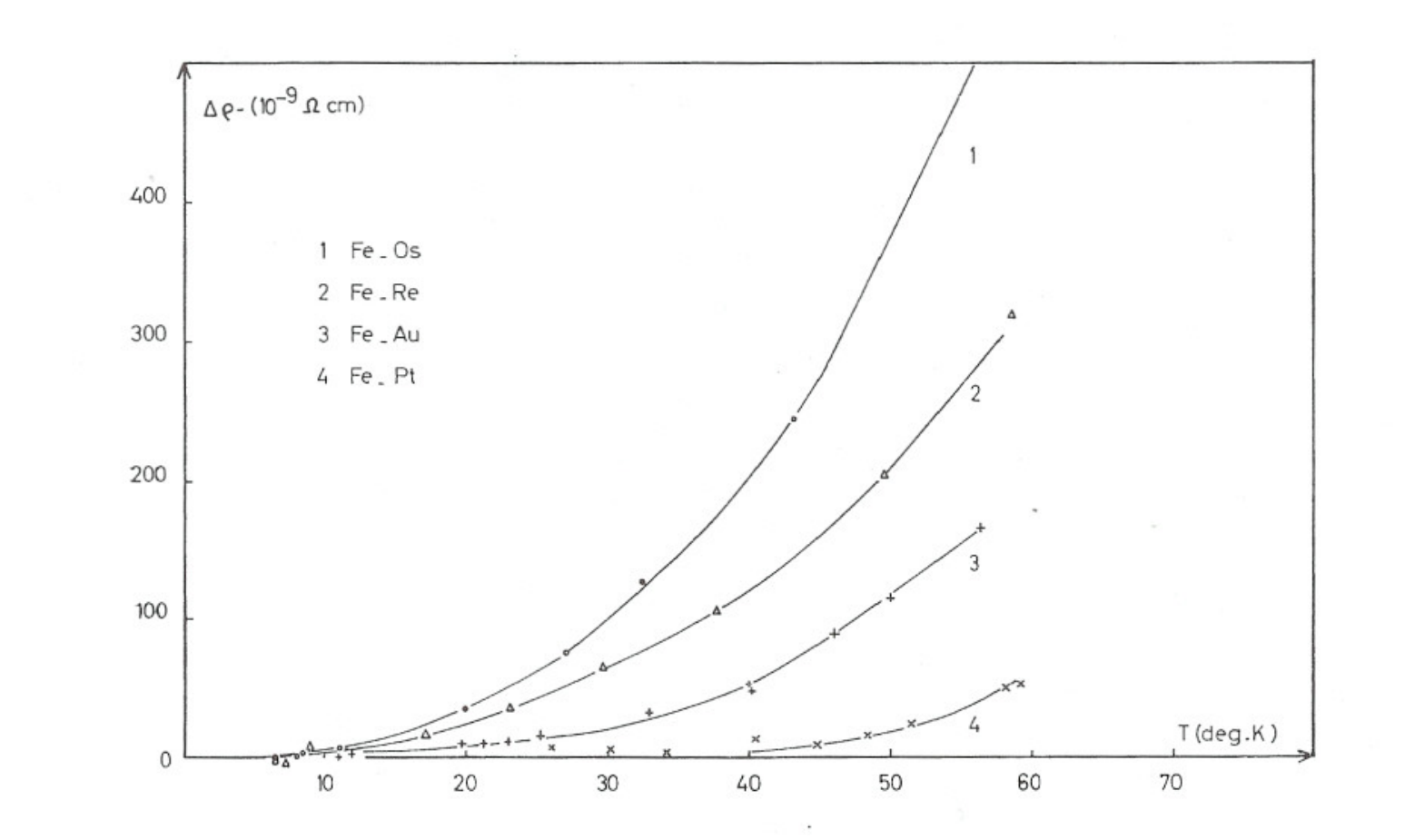}
\caption{The variation of the resistivity deviation $\Delta\rho(T)$ as a function of temperature for a number of alloys for temperatures below $70K$. The alloys are all $3.3(2) wt.\%$.}
\protect\label{fig:1}
\end{figure}

\begin{figure}
\includegraphics[width=4in]{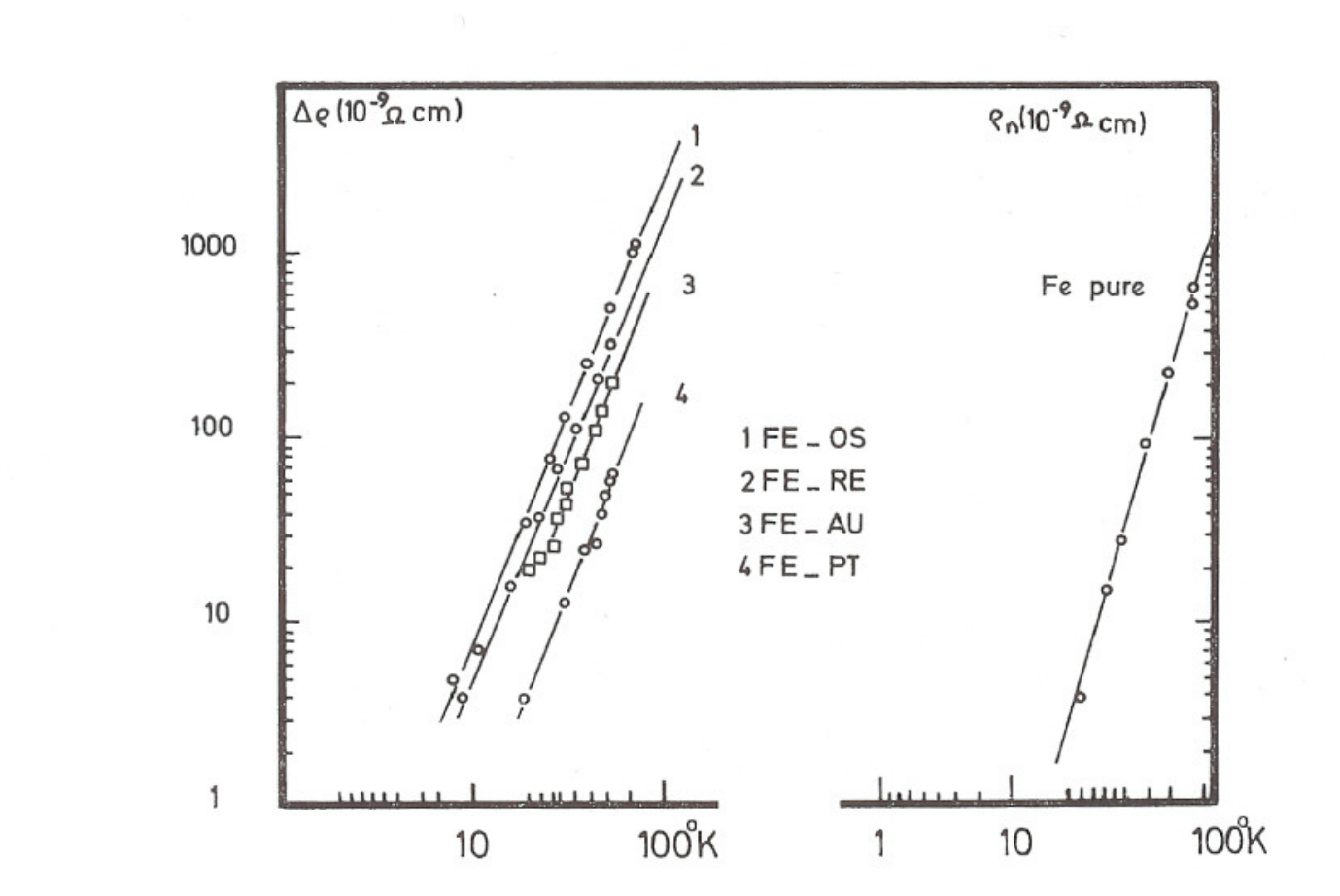}
\caption{log-log plots of the data in (1), plus the resistivity of pure $Fe$ as a function of temperature.}
\protect\label{fig:2}
\end{figure}

\begin{figure}
\includegraphics[width=4in]{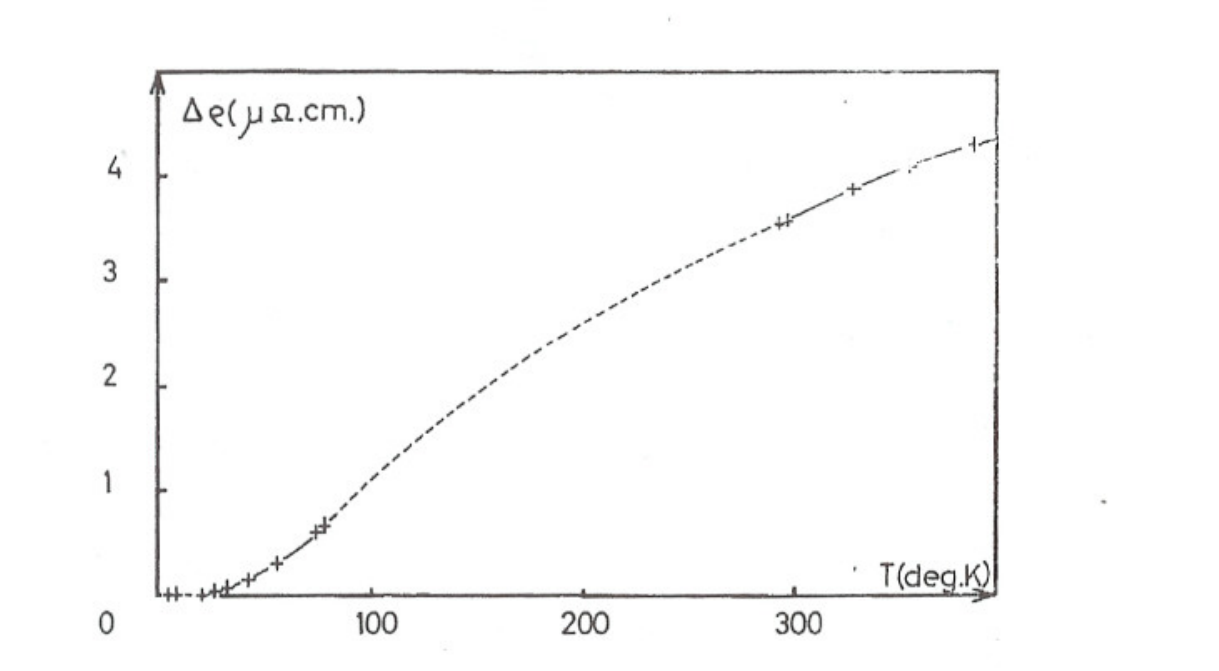}
\caption{An example of the behaviour of the resistivity deviation $\Delta\rho(T)$
over the range of temperature $0K$ to $400K$. The sample is ${\bf Fe}Os 3.3 wt.\%$.}
\protect\label{fig:3}
\end{figure}

\begin{table}
\caption{The deviations from the sum rules at $4.2K$. $\Delta\rho = \rho_{measured}-\rho_{calc}$. The calculated resistivites are those obtained from the sum rule using the single impurity $4.2 K$ specific resistivities from Table 1. Because of innaccuracy in the concentration determinations the uncertainty in $\rho_{calc}$ (and hence in $\Delta\rho$) is $\pm 1 \mu\Omega cm$.
}
\begin{tabular*}{\columnwidth}{c r c r c r c}
\hline
\hline
Impurity conc. wt $\%$ & & $\rho_{4.2 K}\mu\Omega cm$ & & $\rho_{calc} \mu\Omega cm$ & & $\Delta\rho \mu\Omega cm$\\
Al(3.15),V(1.12) & & 42.9&& 35.3 & & 7.6\\
Al(1.05), Ti(2.15) & & 25.4 & & 19.7 & & 5.7\\
Si(1.95), V(1.13)  & &  29.9  & & 24.5  & & 5.4\\
Si(0.89), Mn(2.05) & & 22.6  & & 13.8 & & 8.8\\
Si(1.71), Al(2.35)  & & 44.7 & &  44.8 & & -0.1\\
V(1.93), Mn(2.25)  & & 5.6 & & 6.3 & & -0.7\\
\hline
\hline
\end{tabular*}
\end{table}

These effects can be satisfactorily explained on the following "two current" model. At low temperatures, electrons with spin-up and spin-­down conduct independently and in parallel. The introduction of an impurity gives residual resistivities $\rho_{\uparrow}(0)$ and $\rho_{\downarrow}(0)$ for the two spin directions. As the two currents add to give the total current, the total resistivity $\rho(0)$ will be given by:
\begin{equation}
1/\rho(0) = 1/\rho_{\uparrow}(0) + 1/\rho_{\downarrow}(0)
\end{equation}
or
\begin{equation}
\rho(0) = \rho_{\uparrow}(0)\rho_{\downarrow}(0)/(\rho_{\uparrow}(0) + \rho_{\downarrow}(0))
\end{equation}

As the temperature is increased there are two effects. First a thermal resistance $\rho_{P}^{'}(T)$  must be added for each spin direction; in general, $\rho_{P\uparrow}^{'}(T)\ne \rho_{P\downarrow}^{'}(T)$  Second, electrons can be flipped by spin waves from one spin direction to another, and electron-electron scattering can
take place between electrons of opposite spin, allowing momentum to be transferred from one spin direction to the other. After a total-­momentum-conserving collision, each electron will (on average) have the mean of the two initial momenta. It is important to differentiate between these processes and those which may involve spin-flip, but do not conserve total electron momentum. These last can be included in the thermal resistances.

We can write for each direction of spin the equation for the conservation of total momentum. The rate of creation of momentum by the accele­ration of electrons of charge $e$ and mass $m$ by the electric field $V$ is balanced by the destruction of momentum by scattering, and by inter­change of momentum between the two spin directions. The rate of loss of momentum is equal to the product of the scattering rate and the average momentum lost in each collision. Thus we have:
\begin{equation}
-(e/m)V= \bar{v}_{\uparrow}/\tau_{\uparrow} + (\bar{v}_{\uparrow}-\bar{v}_{\downarrow})/2\tau_{\uparrow\downarrow}
\end{equation}
\begin{equation}
-(e/m)V= \bar{v}_{\downarrow}/\tau_{\downarrow} + (\bar{v}_{\downarrow}-\bar{v}_{\uparrow})/2\tau_{\uparrow\downarrow}
\end{equation}

Then, if $n$ is the density of conduction electrons of each spin, we can write for each current $i$ and resistivity $\rho$:
$i = ne\bar{v}$ and $\rho = m/(ne^2\tau)$
and so we put:
\begin{equation}
1/\tau_{\uparrow} = (ne^2/m)[\rho_{\uparrow}(0) + \rho_{P\uparrow}^{'}(T)]
\end{equation}
\begin{equation}
1/\tau_{\downarrow} = (ne^2/m)[\rho_{\downarrow}(0) + \rho_{P\downarrow}^{'}(T)]
\end{equation}

The $\rho_{P\uparrow,\downarrow}^{'}(T)$ are the temperature dependent parts of the resistivities in the two spin directions. They need not be exactly equal to the temperature dependent resistivities in the pure metal, $\rho_{P\uparrow,\downarrow}(T)$, but will be related, as the phonon scattering will be similar to that in the pure metal.

Similarly, $\rho_{\uparrow\downarrow}(T) = m/ne^2\tau_{\uparrow\downarrow}$
is the momentum-conserving scattering involving electrons of opposite spins.

Finally, the total resistivity at temperature T is:
\begin{equation}
\rho(T) = V/(i_{\uparrow}+i_{\downarrow})
\end{equation}
This gives from eqns. (6) to (10) :
\begin{equation}
\rho(T)=\frac{[\rho_{\uparrow}(0)+\rho_{P\uparrow}^{'}(T)][\rho_{\downarrow}(0)+\rho_{P\downarrow}^{'}(T)]\\+(1/2)\rho_{\uparrow\downarrow}(T)[\rho_{\uparrow}(0)+\rho_{P\uparrow}^{'}(T)+\rho_{\downarrow}(0)+\rho_{P\downarrow}^{'}(T)]}{[\rho_{\uparrow}(0)+\rho_{P\uparrow}^{'}(T)+\rho_{\downarrow}(0)+\rho_{P\downarrow}^{'}(T)+2\rho_{\uparrow\downarrow}(T)]}
\end{equation}	
At low temperatures both the $\rho_{P}^{'}(T)$ and the $\rho_{\uparrow\downarrow}(T)$ are small. For the simplest case of : $\rho_{P\uparrow}^{'}(T)= \rho_{P\uparrow}(T)=\rho_{P\downarrow}^{'}(T)= \rho_{P\downarrow}(T)$ we get for low $T$
\begin{equation}
\rho(T)=\rho(0) + (1/2)\rho_{P}(T)+(1/2)[\frac{\rho_{\uparrow}(0)-\rho{\downarrow}(0)}{\rho_{\uparrow}(0)+\rho{\downarrow}(0)}]^2[\rho_{P}(T)+\rho_{\uparrow\downarrow}(T)]
\end{equation}
The third term is the deviation from Matthiessen's rule, eqn (1). For the general case the deviation does not separate out quite as simply; however it is always positive. The physical picture is that the electrons in the low resistance spin direction, which carry most of the current at low temperatures, are braked as the temperature rises and they are brought into contact with the slower electrons. If $\rho_{\uparrow\downarrow}(T)$ is due to electron-­electron scattering it will be proportional to $T^2$.

So a temperature deviation would be expected for alloys containing impurities for which $\rho_{\uparrow}(0)/\rho_{\downarrow}(0) \ne 1$, and not for those for which $\rho_{\uparrow}(0)/\rho_{\downarrow}(0) \sim 1$. Rough estimates of the ratios for two sorts of impurities can be made. The transition metal impurities are shielded predomi­nantly by the $3d$ electrons (see Friedel \cite{Friedel1962}); as the shielding effect will be quite different for the two directions of spin, $\rho_{\uparrow}(0)$ and $\rho_{\downarrow}(0)$  will probably be very different. Calculations by Gomes \cite{Gomes1966} give ratios of 5 to 10. For alloys containing $Al$ and $Si$ on the other hand, the shielding is mostly by the $s$ electrons, leaving the $d$ bands unaffected (see Mott \cite{Mott1964}). For these cases then $\rho_{\uparrow}(0)\sim \rho_{\downarrow}(0)$. Thus an {\it a priori} prediction would be that samples containing transition impurities would show strong tempera­ture deviations, but those containing $Al$ or $Si$ would not. This is what is found experimentally.

That the low temperature deviation is proportional to $T^2$ suggests electron-electron scattering is responsible; but other processes, such as spin waves and spin-orbit scattering at the impurities should also be taken into account.

The deviation should saturate at high temperatures; we will then have:
\begin{equation}
\rho(T)= (1/4)[\rho_{\uparrow}(0) +\rho_{\downarrow}(0)+\rho_{P\uparrow}^{'}(T)+\rho_{P\downarrow}^{'}(T)]
\end{equation}
If the $\rho_{P}^{'}(T)$ are similar to the pure metal values $\rho_{P}(T)$ , the "apparent residual resistivity" from high temperature measurements is
\begin{equation}
(1/4)[\rho_{\uparrow}(0) + \rho_{\downarrow}(0)]
\end{equation}
which is always larger than the low temperature residual resistivity,
\begin{equation}
\rho(0) = \rho_{\uparrow}(0)\rho_{\downarrow}(0)/(\rho_{\uparrow}(0) + \rho_{\downarrow}(0))
\end{equation}

Now let us consider the deviations from the residual resistivity sum rule, eqn (2). Suppose that an alloy contains two impurities simultaneously, and that the sum rule for residual resistances holds for each direction of spin individually. The residual resistance will be given by:
\begin{equation}
1/\rho_{1,2} = 1/(\rho_{1\uparrow}+ \rho_{2\uparrow})+ 1/(\rho_{1\downarrow}+ \rho_{2\downarrow})
\end{equation}

Simple algebra shows that $\rho_{1,2}$  is always greater than the sum of the individu al residual resistivities $\rho_{1}+\rho_{2}$ which would be measured if the impurities were present, one at a time, with the same concentrations, unless $\rho_{1\uparrow}/\rho_{1\downarrow} = \rho_{2\uparrow}/\rho_{2\downarrow}$. Thus the deviations actually observed from the residual resistivity sum rule would be expected if $Mn$, $Cr$ and $V$ impurities all had similar $\rho_{\uparrow}/\rho_{\downarrow}$  ratios, but these ratios were very different from those of $Al$ and $Si$ impurities. These results are consistent with the estimates of the $\rho_{\uparrow}/\rho_{\downarrow}$  ratios given earlier and with the results for the temperature deviations.

A first remark to make in conclusion is that the impurity resistivity values in ferromagnetic metals deduced from measurements at room temperature are of only very qualitative significance as physical parameters as they generally represent neither the real $0K$ residual resistivity nor the high temperature limit.

There are numerous measurements of interest which can test and amplify this model. Detailed measurements over a range of relative concentrations for alloys containing two impurities simultaneously should give actual values of $\rho_{\uparrow}/\rho_{\downarrow}$   ratios, which will in turn give information on the electronic structure of the alloys. Temperature and concentration dependances for alloys containing single impurities could shed light on the electron-electron scattering, spin wave scattering, and spin-orbit scattering effects. There are also immediate extensions to the other transport properties of ferromagnetic metals.

\begin{acknowledgements}
We should like to thank Professor P\'erio who gave us the opportunity of performing a number of experiments in his laboratory and Mr. Lauriat for his help with the measurements.
\end{acknowledgements}

\end{document}